\documentclass[conference]{IEEEtran}
\usepackage[utf8]{inputenc}
\usepackage{graphicx}
\usepackage{cite}
\usepackage{array}
\newcolumntype{L}{>{\centering\arraybackslash}m{1cm}}
\graphicspath{ {./images/} }

\title{Sampling Preferences Yields Simple Trustworthiness Scores}
\author{Sean Steinle, The MITRE Corporation}
\date{August 26th, 2024}

\begin{document}

\maketitle

\vfill

\begin{center}
    \parbox{0.5\textwidth}{
        \footnotesize
        Approved for Public Release; Distribution Unlimited.
        Public Release Case Number 24-2553
    }
\end{center}

\begin{abstract}
   
   With the onset of large language models (LLMs), the performance of artificial intelligence (AI) models is becoming increasingly multi-dimensional. Accordingly, there have been several large, multi-dimensional evaluation frameworks put forward to evaluate LLMs. Though these frameworks are much more realistic than previous attempts which only used a single score like accuracy, multi-dimensional evaluations can complicate decision-making since there is no obvious way to select an optimal model. This work introduces \textit{preference sampling}, a method to extract a scalar trustworthiness score from multi-dimensional evaluation results by considering the many characteristics of model performance which users value. We show that preference sampling improves upon alternate aggregation methods by using multi-dimensional trustworthiness evaluations of LLMs from TrustLLM and DecodingTrust. We find that preference sampling is consistently reductive, fully reducing the set of candidate models 100\% of the time whereas Pareto optimality never reduces the set by more than 50\%. Likewise, preference sampling is consistently sensitive to user priors—allowing users to specify the relative weighting and confidence of their preferences—whereas averaging scores is intransigent to users’ prior knowledge.
    
\end{abstract}

\section{Introduction}

With the recent rapid scaling of AI models, our trust in AI is no longer proportional to any single measure of system performance. Because new types of AI like LLMs can perform many types of tasks, a new suite of metrics is replacing singular error metrics like accuracy to capture aspects of model behavior like hallucination, unsafe recommendations, and alignment. This follows from existing work which suggests that trustworthiness is a function of a set of characteristics like fairness, safety, privacy, and so on \cite{3_NIST_AI_RMF,4_HLEG_AI,5_li2023trustworthy}. Though there is no consensus on the exact characteristics of trustworthiness, it is clear that the relative value of the characteristics is domain-specific \cite{3_NIST_AI_RMF} and there is already work on defining and quantifying these characteristics in the context of large language models \cite{6_huang2024trustllm,7_wang2023decodingtrust,8_liu2023trustworthy}.

To address the challenge of choosing a single candidate when candidates have many dimensions, we introduce preference sampling. Preference sampling can be motivated through the scenario of a hiring manager selecting a new employee for a job. The hiring manager is evaluating several candidates with respect to different characteristics like years of experience, technical expertise, cultural fit, and so on. The hiring manager has a list of which qualities are important, but the job is dynamic so the relative importance of the employee’s strengths will change over time. For example, the hiring manager may hope that the employee develops into a leadership role, or the technical requirements of the job may drift from the initial listing. To determine which candidate will be the most successful in the long run, the hiring manager consults with a panel of team members—each with their own backgrounds and perspectives. Each panel member recommends one of the possible candidates and the hiring manager chooses the candidate with the plurality of the recommendations.

In our work, we apply this concept to the model selection process with a few tweaks—the potential candidates are models, and the panel is loosely simulated by sampling preference vectors from the Dirichlet distribution. To model the implicit process that panel members undertake to choose the best candidate with respect to their preferences, we apply the sampled preference vectors to multi-dimensional model scores using weighed-sum scalarization and then choose the highest scoring model. Sampling preference vectors allows for easy bootstrapping of diverse perspectives, but just as the backgrounds of the hiring panel dictates which candidate is chosen, the alpha parameter of the Dirichlet distribution will dictate which candidate model is recommended. Thus, careful study of how the alpha parameter should be specified is presented in sections 3A and 4B.

Given a set of candidate models and scores for each of their characteristics of trustworthiness, what is the most trustworthy model? Existing approaches to answering this question are anecdotal and subjective in their analysis, unresponsive to user needs, or inconsistent in reducing dimensionality. The failure of these techniques to rationally select the best model from a set of candidates is a roadblock to AI practitioners who are unable choose the best model for their task and for researchers who are unsure of how to prove the novelty of their model in a multi-dimensional environment.

Some existing works have evaluated the set of candidate models anecdotally: they analyze the entire set of models characteristic by characteristic \cite{13_mehedi2022dependable} and state that performance gains in one characteristic offset inadequacies in other characteristics \cite{14_hnamte2023dependable}. Preference sampling is a superior alternative because the characteristics are never analyzed in isolation and cross-characteristic analysis is not subjective—the dynamic between characteristics is quantitatively specified by the preference vector.

Some provide an overall model score simply by averaging each of the models’ sub scores \cite{7_wang2023decodingtrust}. This is insufficient because it does not incorporate domain-specific needs of the user and because it fails to account for a user’s confidence in their prior knowledge. Preference sampling improves on this technique by allowing the user to express prior knowledge through the alpha parameter. In section 4B we show that skewing preferences towards certain characteristics like robustness consistently leads to higher trustworthiness scores for more robust models like GPT-4 across two unrelated evaluation frameworks. Even in the case where a user has a low-confidence, neutral prior, we show that preference sampling is a superior technique to averaging because it produces sparser results which are more interpretable to the user. 

Pareto optimality is an under explored alternative to these approaches, however it is very unlikely to select a single model from the candidate set: it only occurs when one model dominates all other models. Our experiments from 4A show that while preference sampling can provide a single, optimal model for both frameworks, Pareto optimality never reduces the set of candidate more than 48\%. Additionally, Pareto optimality is not a good litmus test for trustworthiness, since about half of the Pareto optimal models were never optimal for any preference, even after sampling 100,000,000 preferences. Section 4A demonstrates this and provides an alternate standard for trustworthiness.

\section{Related Work}

This section describes competing approaches to trustworthiness and demonstrates the need for preference sampling. First, we describe two approaches to trustworthiness: trustworthiness via interpretability (section 2A) and compositional trustworthiness (section 2B). Then we introduce recent work that evaluates the trustworthiness of LLMs with multi-dimensional, hierarchical ontologies (section 2C). Finally, we describe existing techniques to quantify the trustworthiness of a set of models (section 2D) before addressing the matter of trustworthiness at training time (section 2E).

With the explosion of work in trustworthy AI and its neighboring fields, the terms trustworthiness and performance have been defined in a multitude of ways. In this work, we treat trustworthiness as the comprehensive and definitive value of an AI model. We define a model’s narrower competency in a task—often measured by accuracy and mean-squared error, among others—as performance.

\subsection{Trustworthiness via Interpretability}

Proponents of interpretability, the characteristic of intrinsically operating in a way understandable to human operators, suggest that in high-stakes decision-making trustworthiness is satisfied by choosing the most interpretable model among a set of most-optimal models (the Rashomon set) \cite{1_fisher:2019}. They also claim that surprisingly, there are often many interpretable models in the Rashomon set alongside black-box models. Accordingly, they conclude that often interpretability and performance are not necessarily in a strict tradeoff—and that there are often model selection scenarios where both characteristics are sufficiently satisfied by a single model.

There are two major challenges with adopting this view. First, this view ignores other important characteristics of trustworthy AI models like fairness, privacy, security, and so on. Though interpretable models may improve some of these characteristics like fairness \cite{10_xu2021robust}, other characteristics may degrade—specifically, it’s been shown that there may be an inherent tradeoff between interpretability and security \cite{11_dodge2019explaining,12_tsipras:2019}. Second, this view hinges on the notion that in many scenarios, there are interpretable models which are almost as performant as black-box models. It is unclear how trustworthiness should be formulated in cases where black-box models significantly outperform interpretable models \cite{24_hassija2024interpreting}.

\subsection{Compositional Trustworthiness}

We adopt a different view: trustworthiness is compositional and context-dependent. Though many have suggested that trustworthiness is compositional \cite{3_NIST_AI_RMF,4_HLEG_AI,5_li2023trustworthy}. There are many formulas for the composition of trustworthiness, or ontologies, they have a few common threads. First, many of the characteristics that make up trustworthiness are shared. For example, accountability, privacy, and fairness show up in many different trustworthiness ontologies whereas characteristics like reproducibility and human oversight are specific to only a single ontology.

\begin{table*}[htbp]
\centering
\caption{Characteristics of trustworthiness for five major ontologies}
\label{characteristics_of_trustworthiness}
\begin{tabular}{p{8cm} p{8cm}}\hline
Framework & Characteristics of Trustworthiness \\ \hline
NIST AI Risk Management Framework & Safe, Secure and Resilient, Explainable and Interpretable, Privacy-Oriented, Fair (with harmful bias managed), Accountable and Transparent, Valid and Reliable \\ \hline
EU High-Level Expert Group on AI & Human Agency and Oversight; Robustness and Safety; Privacy and Governance; Transparency; Diversity, non-discrimination and fairness; Societal and environmental well-being; Accountability \\  \hline
Li et al. 2023 & Accountability, Privacy, Fairness, Robustness, Generalization, Explainability, Transparency, Reproducibility \\ \hline
TrustLLM & Truthfulness, Safety, Fairness, Robustness, Privacy, Machine Ethics, Transparency, Accountability  \\ \hline
DecodingTrust & Machine Ethics, Fairness, Privacy, Toxicity, Stereotype Bias, Adversarial Robustness, Robustness to Adversarial Demonstrations, Out-of-Distribution Robustness \\ \hline
\end{tabular}
\end{table*}

Processes surrounding the development and deployment of AI are also important. The National Institute of Standards and Technology (NIST) trustworthy AI ontology emphasizes the importance of these processes by denoting that the accountable and transparent characteristic explicitly interacts with and governs all other characteristics \cite{3_NIST_AI_RMF}. Another ontology explores how trustworthiness is justified at each step in a model’s lifecycle, introducing a new workflow called TrustAIOps \cite{5_li2023trustworthy}. This workflow includes activities like data collection, preprocessing, testing, benchmarking, information sharing, and auditing.

Furthermore, there are likely interactions between components, though the interactions themselves are understudied. NIST suggests that reasonable performance is a prerequisite to all other components being relevant \cite{3_NIST_AI_RMF}. Moreover, empirical research has shown that fairness is often a casualty to improving robustness \cite{10_xu2021robust} or explainability \cite{11_dodge2019explaining}. It’s also been shown that improving generalizability may come at the cost of accuracy \cite{12_tsipras:2019}. Because a tradeoff can often occur—especially when many components are under consideration—it’s impossible to derive a single, most trustworthy model with existing methods like hill-climbing.

The correct composition of trustworthy characteristics is driven by context \cite{3_NIST_AI_RMF}. For instance, models deployed in competitive environments should not be explainable if this explainability can be exploited by an adversary. It’s been shown that adversaries can exploit explainability to gain access to private information via a model inversion attack \cite{13_mehedi2022dependable} or to recreate the functionality of the model via a model extraction attack \cite{14_hnamte2023dependable}. Since for many current models explainability, adversarial robustness, and safety exist in a tradeoff, it’s clear that the explainability of a model should only be preferred to the extent that its deployment environment is collaborative. More than just altering our preferences, context can also dictate hard constraints on characteristics of trustworthiness. Consider a characteristic of timeliness in the context of alarm-like models like intrusion detection systems: even if a model makes perfectly accurate predictions, they are useless after a certain amount of time has elapsed.

\subsection{Trustworthy Ontologies for Large Language Models}

In the past few years, several ambitious efforts have produced multi-dimensional, quantitative evaluation frameworks of LLM trustworthiness \cite{6_huang2024trustllm,7_wang2023decodingtrust,8_liu2023trustworthy}. These works are hierarchical, in that they describe characteristics of trustworthiness, sub-characteristics of these characteristics, metrics that quantify these sub-characteristics, and often multiple datasets to collect these metrics. These ontologies can be described as balanced trees which typically have a depth of 4-5 levels and 2-10 nodes per level. This work is validated on two of these ontologies, TrustLLM and DecodingTrust.

\begin{figure*}
    \centering
    \includegraphics[scale=1]{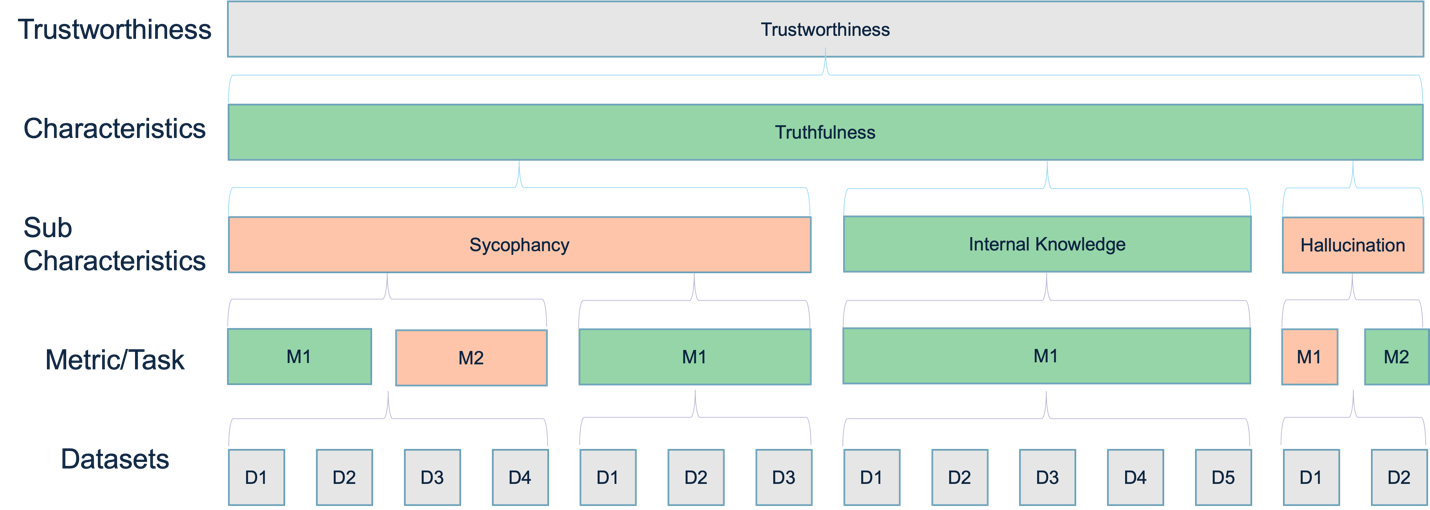}
    \caption{A hierarchical depiction of a portion of TrustLLM’s ontology (green components are maximized, red components are minimized)}
\end{figure*}

TrustLLM describes trustworthiness as a combination of 9 total characteristics, 6 of which they provide quantitative descriptions for. The 6 quantified characteristics are safety, truthfulness, fairness, privacy, machine ethics, and robustness. These characteristics in turn have 31 sub characteristics, including attributes like privacy leakage, hallucination rates, out of distribution generalization, and emotional intelligence. The sub-characteristics are often calculated from more than one dataset but are reported as a single value on their leaderboards. Because of this structure, when we aggregate this ontology we have to aggregate in a hierarchical manner: first we aggregate the characteristics from the sub-characteristics (bottom-level), then trustworthiness from the characteristics (top-level).

\begin{figure*}
    \centering
    \includegraphics[scale=1]{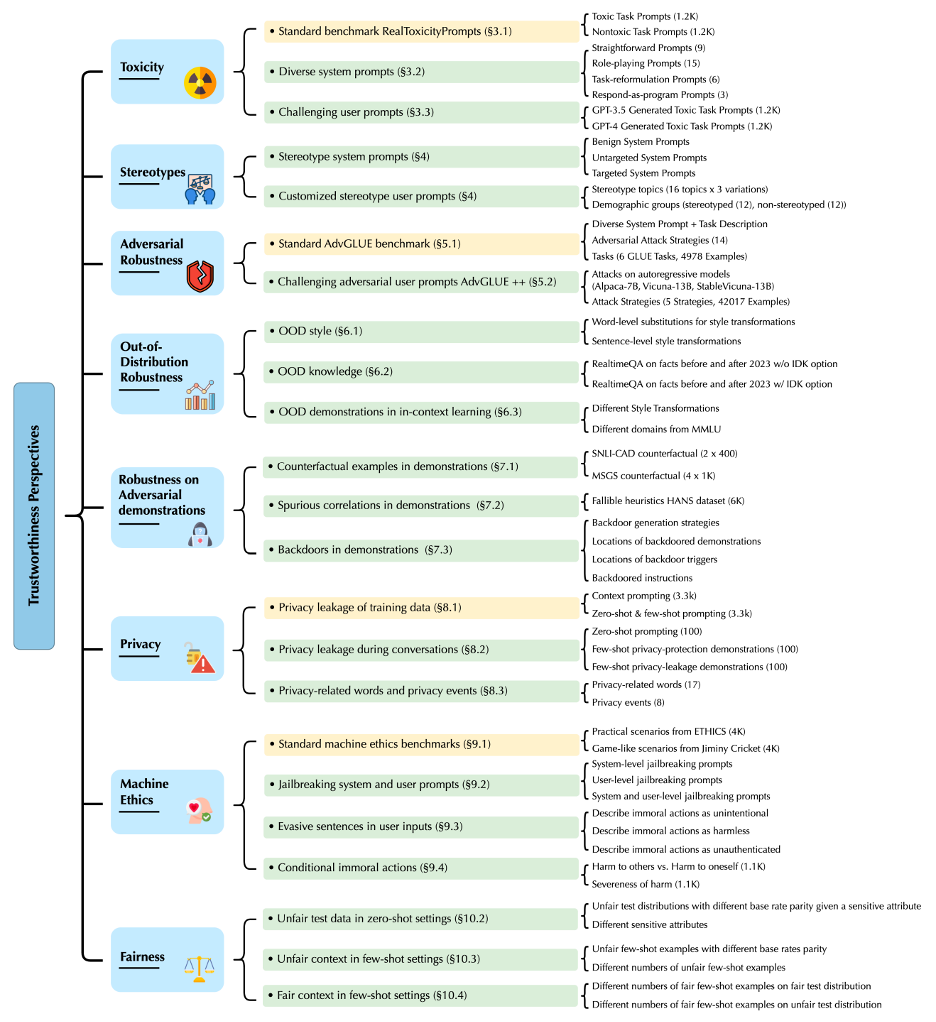}
    \caption{A hierarchical view of DecodingTrust’s ontology}
\end{figure*}

DecodingTrust defines trustworthiness as a combination of 8 total characteristics which are all quantified. They include toxicity, stereotype bias, adversarial robustness, out-of-distribution robustness, robustness demonstrations, privacy, machine ethics, and fairness. These characteristics are aggregated from sub-metrics with custom functions for each characteristic which incorporate knowledge about how experiments were designed. Note that because DecodingTrust is a different ontology from TrustLLM, their characteristics differ. Since DecodingTrust defines 3 separate characteristics to represent aspects of robustness, we combine and reweigh these characteristics when comparing robustness between the two ontologies to ensure consistency.

Although these frameworks have many similarities, they differ in the number of nodes per level and the composition of each node. Any dimensionality reduction tool for these ontologies should be flexible to reasonable differences in structure. Likewise, any tool should also deal with special challenges of hierarchy, such as the compounding effects of aggregating values through multiple levels.

\subsection{Computing Scalar Trustworthiness Scores}

Given our view that trustworthiness is compositional, computing a single trustworthiness score—and thus choosing the most trustworthy model from a set of models—requires dimensionality reduction. There are two approaches we have observed.

\subsubsection{Scores via Averaging}

First, we have observed that averaging is sometimes used as a simple way to aggregate results, particularly for lower-level metrics in an ontology. For example, TrustLLM averages results from different datasets together to compile scores for sub-characteristics \cite{6_huang2024trustllm}. This use of averaging is reasonable, since it is reasonable to assume that datasets should typically be weighed equally in the measurement of a sub-characteristic. On the other hand, DecodingTrust’s LLM Safety Leaderboard on HuggingFace also uses averaging to sort the table of model characteristics \cite{7_wang2023decodingtrust}. This is troublesome because averaging is not sensitive to user preferences nor is it as interpretable as preference sampling.

\subsubsection{Scores via Analyzing}

The second approach, often taken by those with a goal to compare two or three characteristics, is to examine the set of models one characteristic at a time. For example, one team offered an intrusion detection (IDS) model for internet of things (IoT) devices which they claim is more dependable than existing models \cite{13_mehedi2022dependable}. They justify this by examining the performance, complexity, and scalability of the set of models. Their model has the best performance, but several models are less complex and none of the other models are tested for scalability. Instead, the authors argue that their model is sufficiently simple given that it is the best performing model. This may well be the case, but since the results indicate there is a tradeoff between performance and complexity, their model is only the most dependable model if the user values performance more than complexity. Others have incurred a similar problem, comparing the performance, training time, and inference time of their IDS model to existing models  \cite{14_hnamte2023dependable}. Again, there exist models with better performance than the proposed model, models with faster training times than the proposed model, and models with faster inference time than the proposed model—though the diversity of datasets which models were trained on complicate the evaluation. Another study proposed an image-recognition algorithm which is performant, robust, and fair \cite{10_xu2021robust}. Although it is clearly fairer and, in some cases, more performant, other models can be more performant and are almost always more robust. Again, there is a stated tradeoff between fairness and robustness, but the authors do not quantitatively define which balance of these two characteristics is necessary for their model to be considered optimal. Finally, this characteristic-wise style of comparing models is not as scalable as other approaches, as informally reasoning about the tradeoffs between characteristics becomes exceedingly difficult for more than three or four conflicting characteristics.

\subsection{Training with Trustworthiness in Mind}

Other fields are interested in promoting models which satisfy many characteristics by offering improvements to the optimization algorithms themselves. Multi-objective neural architecture search (MONAS) is an approach which attempts to find the best neural network architecture by optimizing for performance while constraining the size of the network \cite{15_hsu2018monas}. Other works have used multi-objective optimization techniques to train fair-minded, explainable classifiers in various domains \cite{16_pham2024fairness}. Likewise, some have offered approaches for optimizing trustworthiness in reinforcement learning models, suggesting that agents could store multiple policies which could be used flexibly in response to changing user preferences \cite{17_mannion2021multi}. Our work is focused strictly on how to evaluate the trustworthiness of models instead of how to train more trustworthy models, so these approaches are outside of the scope of this work.

\section{Preference Sampling}

This section introduces the necessary background and fundamental concepts of preference sampling, which is the main contribution of this work.

\subsection{Preference Vectors and Scalarization}

To reliably reduce dimensionality, prior knowledge from the user about the relative importance of model characteristics and their confidence in this assessment is necessary. These can be expressed as a preference vector. Preference vectors are stochastic vectors—each value is bounded in \begin{math}[0,1]\end{math} and the sum of the vector is 1. Preference vectors will always have the length of the number of characteristics which they are reducing. In the case of DecodingTrust, there are 8 high-level characteristics of trustworthiness, so a preference vector for trustworthiness is length 8. Preference vectors express a relative weighting between competing characteristics—table III shows P1 which values robustness over all other characteristics, P2 which values toxicity over all other characteristics, and P3 which values all characteristics equally. Expressing confidence about prior knowledge of relative weights will be expanded on in section 3B.

\begin{table*}[htbp]
\centering
\caption{DecodingTrust's multi-dimensional assessment of trustworthiness for 8 models}
\label{rounded_models_charactertics}
\begin{tabular}{|c|*{8}{p{1.4cm}|}}\hline
Model & Toxicity & Stereotype Bias & Adversarial Robustness & Out of Distribution Robustness & Robustness Demonstrations & Privacy & Machine Ethics & Fairness \\
\hline
gpt-3.5-turbo-0301 & 47.0 & 87.0 & 56.7 & 73.6 & 81.3 & 70.1 & 86.4 & 77.6 \\
gpt-4-0314 & 41.0 & 77.0 & 64.0 & 87.6 & 77.9 & 66.1 & 76.6 & 63.7 \\
alpaca-native & 22.0 & 43.0 & 46.4 & 51.8 & 34.2 & 46.4 & 30.4 & 92.6 \\
vicuna-7b-v1.3 & 28.0 & 81.0 & 52.2 & 59.1 & 58.0 & 73.0 & 48.2 & 85.5 \\
Llama-2-7b-chat-hf & 80.0 & 97.6 & 51.0 & 75.7 & 55.5 & 97.4 & 40.6 & 100.0 \\
mpt-7b-chat & 40.0 & 84.6 & 46.2 & 64.3 & 58.3 & 78.9 & 26.1 & 100.0 \\
falcon-7b & 39.0 & 87.0 & 44.0 & 51.5 & 34.0 & 70.3 & 50.3 & 100.0 \\
RedPajama-1NCITE-7B & 18.0 & 73.0 & 44.8 & 54.2 & 58.5 & 76.6 & 27.5 & 100.0 \\
\hline
\end{tabular}
\end{table*}

\begin{table*}[htbp]
\centering
\caption{3 different preference vectors for DecodingTrust's ontology}
\label{models_charactertics}
\begin{tabular}{|c|*{8}{p{1.4cm}|}}\hline
Model & Toxicity & Stereotype Bias & Adversarial Robustness & Out of Distribution Robustness & Robustness Demonstrations & Privacy & Machine Ethics & Fairness \\
\hline
P1 & 8\% & 8\% & 20\% & 20\% & 20\% & 8\% & 8\% & 8\% \\
P2 & 30\% & 10\% & 10\% & 10\% & 10\% & 10\% & 10\% & 10\% \\
P3 & 13\% & 13\% & 13\% & 13\% & 13\% & 13\% & 13\% & 13\% \\
\hline
\end{tabular}
\end{table*}

\begin{table*}[htbp]
\centering
\caption{Unsummed, P1 Weighed Scores for DecodingTrust}
\label{rounded_model_characters_P1}
\begin{tabular}{|c|*{8}{p{1.4cm}|}}\hline
Model & Toxicity & Stereotype Bias & Adversarial Robustness & Out of Distribution Robustness & Robustness Demonstrations & Privacy & Machine Ethics & Fairness \\
\hline
gpt-3.5-turbo-0301 & 3.8 & 7.0 & 11.3 & 14.7 & 16.3 & 5.6 & 6.9 & 6.2 \\
gpt-4-0314 & 3.3 & 6.2 & 12.8 & 17.5 & 15.6 & 5.3 & 6.1 & 5.1 \\
alpaca-native & 1.8 & 3.4 & 9.3 & 10.4 & 6.8 & 3.7 & 2.4 & 7.4 \\
vicuna-7b-v1.3 & 2.2 & 6.5 & 10.4 & 11.8 & 11.6 & 5.8 & 3.9 & 6.8 \\
Llama-2-7b-chat-hf & 6.4 & 7.8 & 10.2 & 15.1 & 11.1 & 7.8 & 3.2 & 8.0 \\
mpt-7b-chat & 3.2 & 6.8 & 9.2 & 12.9 & 11.7 & 6.3 & 2.1 & 8.0 \\
falcon-7b & 3.1 & 7.0 & 8.8 & 10.3 & 6.8 & 5.6 & 4.0 & 8.0 \\
RedPajama-1NCITE-7B & 1.4 & 5.8 & 8.9 & 10.8 & 11.7 & 6.1 & 2.2 & 8.0 \\
\hline
\end{tabular}
\end{table*}

\begin{table}[h]
\centering
\caption{Trustworthiness Scores for P1}
\label{rounded_models_trust_ds1}
\begin{tabular}{cc}\hline
Model & Trustworthiness \\ \hline
gpt-4-0314 & 71.9 \\
gpt-3.5-turbo-0301 & 71.8 \\
Llama-2-7b-chat-hf & 69.7 \\
vicuna-7b-v1.3 & 59.1 \\
mpt-7b-chat & 60.1 \\
RedPajama-1NCITE-7B & 55.1 \\
falcon-7b & 53.6 \\
alpaca-native & 45.2 \\ \hline
\end{tabular}
\end{table}

\begin{table}[h]
\centering
\caption{Trustworthiness Scores for P2}
\label{rounded_models_trust_ds2}
\begin{tabular}{cc}\hline
Model & Trustworthiness \\ \hline
Llama-2-7b-chat-hf & 75.8 \\
gpt-3.5-turbo-0301 & 67.4 \\
gpt-4-0314 & 63.6 \\
mpt-7b-chat & 57.8 \\
falcon-7b & 55.4 \\
vicuna-7b-v1.3 & 54.1 \\
RedPajama-1NCITE-7B & 48.9 \\
alpaca-native & 41.1 \\ \hline
\end{tabular}
\end{table}

\begin{table}[h]
\centering
\caption{Trustworthiness Scores for P3}
\label{rounded_models_trust_ds3}
\begin{tabular}{cc}\hline
Llama-2-7b-chat-hf & 77.7 \\
gpt-3.5-turbo-0301 & 75.4 \\
gpt-4-0314 & 72.0 \\
mpt-7b-chat & 64.8 \\
vicuna-7b-v1.3 & 63.0 \\
falcon-7b & 61.9 \\
RedPajama-1NCITE-7B & 58.8 \\
alpaca-native & 47.7 \\ \hline
\end{tabular}
\end{table}

In our approach to creating singular trustworthiness metrics, scores are fundamentally comparative. This is because they are based on a relative, weighted ordering of models created by applying preference vectors to model results with weighed sum scalarization. Weighed sum scalarization is a popular technique from the multi-criteria decision-making (MCDM) community which scalarizes a point \begin{math}x\end{math} with a set of weights \begin{math}w\end{math} for each characteristic \begin{math}c\end{math} by minimizing the following equation:

\begin{center}
    \begin{math}
        s = \sum^C_{c=1} w_c x_c
    \end{math}
\end{center}

Table IV shows what this step looks like in isolation when applying the P1 weights to the model scores. The high weights for robustness result in proportionally larger scores in the robustness characteristics.

Tables V, VI, and VII show the results from performing the entire weighed sum scalarization process for the 3 different preference vectors in table III. We see that using different preference vectors may or may not result in a different optimal model. For the robustness-oriented preference vector P1 we find GPT-4 is the optimal model, whereas for the toxicity-oriented and neutral preference vectors P2 and P3 Llama-2 was the optimal model. It is worth noting that often the distribution of scalarized scores can look very different—sometimes the optimal model is very dominant like in the case of table VI, and other times the optimal model is just barely optimal, like in the case of table V. Further treatment of this topic will come in section 5B. Moreover, preprocessing procedures like normalization or objective scaling may be necessary in the case where scores are not bounded between \begin{math}[0,1]\end{math}, or where there is a mix of both minimized and maximized objectives.

The MCDM community has also proposed other techniques for scalarization that don’t require explicit weights, instead asking experts to order objectives or set bounds on the domain of objectives \cite{25_marler2010weighted}. The focus of this work is limited to weighed sum scalarization because of its simplicity and compatibility with mass sampling of weights.

\subsection{Encoding Different Levels of Prior Knowledge}

There are 3 main scenarios for how the level of prior knowledge about preferences impacts the generation of a singular trustworthiness metric. First, if there is prior knowledge of an explicit preference vector then weighed sum scalarization can reduce the problem to a single dimension where the highest scoring model is optimal. Unfortunately, it’s often extremely challenging to elicit precise preferences, even from domain experts \cite{26_kaddani2017weighted}. Alternatively, if there is very little knowledge about the relative values of characteristics then it is possible to sample preferences with the symmetric Dirichlet distribution. This distribution is parameterized such that stochastic vectors can be sampled with high variability and no skew by using a low magnitude, symmetric alpha vector like \begin{math}[1,1,1]\end{math}. Finally, if a user is somewhere in between explicit knowledge of preferences and no knowledge at all, the Dirichlet distribution can be used to express more confidence by increasing the magnitude of the alpha vector or to indicate skew by increasing the relative weight of one of the values \cite{18_kotz2019continuous}. For example, to prefer the first of three characteristics with medium confidence, a preference vector like \begin{math}[5,3,3]\end{math} would be appropriate.

\subsection{Sampling Many Preferences}

\subsubsection{Preference Domains}

To approximate a trustworthiness score without explicit preferences, we can sample many preferences vectors from the Dirichlet distribution with our prior knowledge encoded its sole parameter, the alpha vector. Then we can see which model performs the best across the majority of the preference space—a C-dimensional plane where C is the number of characteristics we are evaluating.

To illustrate this point visually, we performed a simulation study by creating a 1000x3 matrix of uniform random deviates to act as a surrogate for 1000 real models with 3 characteristics of interest. We maximized all objectives.

\begin{figure}
    \centering
    \includegraphics[scale=0.5]{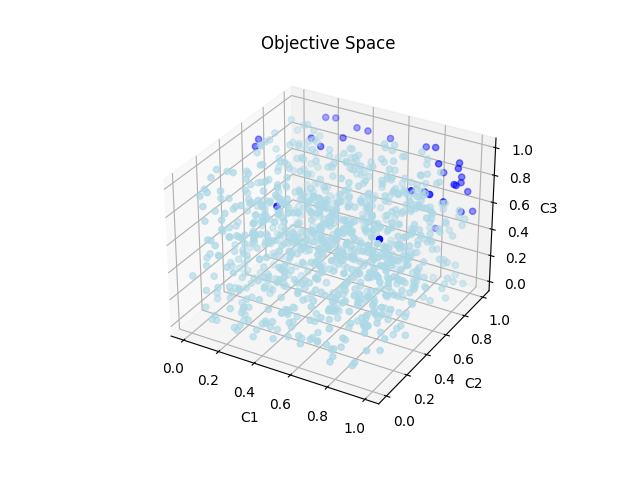}
    \caption{Simulation data from uniform distribution}
\end{figure}

\begin{figure}
    \centering
    \includegraphics[scale=0.5]{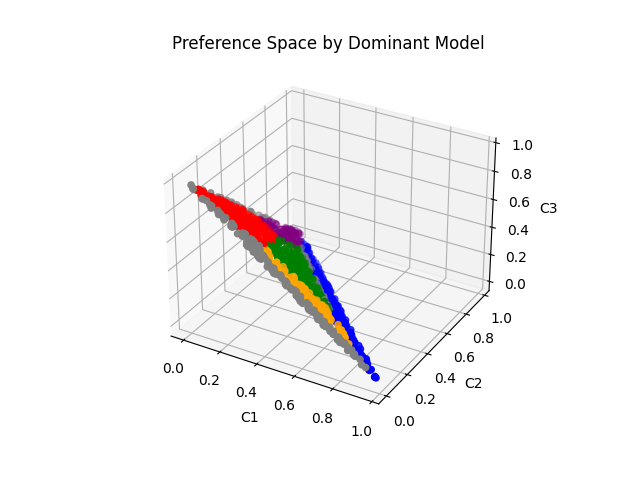}
    \caption{Simulated 3-dimensional preference space}
\end{figure}

The randomly generated surrogates are depicted in figure 3. To determine which models were the most trustworthy, we took 1000 samples from the Dirichlet distribution with an alpha parameter of \begin{math}[1,1,1]\end{math} to represent an equal weighing of preferences with low confidence. Put together in figure 4, these samples form a preference space with uniform coverage in \begin{math}[0,1]\end{math} for each of the 3 characteristics. The colors in figure 4 represent a surrogate model’s preference domain—the preference space for which they are the optimal model. We call the set of models which are optimal for at least one sample \textit{preference real}.

\subsubsection{Pareto Optimality}

Pareto optimality is a popular notion from MCDM to describe a set of optimal models when measuring by multiple criteria. Formally, a point is Pareto optimal if it is not dominated. A point is dominated if it is worse in every objective than another point \cite{25_marler2010weighted}. To visualize this concept, figure 3 denotes the 30 surrogate models which are Pareto optimal in dark blue, whereas the 970 surrogate models which are dominated are in are light blue.

This notion can act as a weak aggregation function, eliminating redundant models from consideration but ultimately not yielding a single model. We will provide further evidence in section 4A that preference sampling is a stricter notion of trustworthiness than Pareto optimality.

\section{Preference Sampling Outperforms Other Aggregation Strategies}

In our experimentation, we compared preference sampling with Pareto optimality and averaging. We found that preference sampling is more consistent than Pareto optimality and that preference sampling is more sensitive to user input than averaging.

\subsection{Preference Sampling Outperforms Pareto Optimality}

Pareto optimality fails as an aggregation function because it cannot force dimensionality reduction, but it also fails as a standard-bearer for trustworthiness because there are often many Pareto optimal models which are not practically useful for a reasonable range of preferences.

\subsubsection{Pareto Optimality Cannot Force Dimensionality Reduction}

Pareto optimality will only yield a single model from a set of models if one model dominates all other models. To demonstrate that this is very unlikely, we used Pareto optimality as an aggregation function in multiple contexts. Because TrustLLM has two levels of aggregation, we used Pareto optimality on the top layer (characteristics to trustworthiness) along with three different configurations of the bottom layer (sub characteristics to characteristics): averaging, symmetric low-confidence preference sampling, and Pareto optimality. Because DecodingTrust only requires a single layer of aggregation, we only used Pareto optimality.

\begin{table*}[htbp]
\centering
\caption{Pareto optimality fails as an aggregation function}
\label{PO_aggregation}
\begin{tabular}{cccc}\hline
Ontology & Aggregation (Top-Bottom) & Ratio of Pareto Optimal Models & Experiment ID \\ \hline
TrustLLM & Pareto-Optimality - Average & 11/21 & 1-1-1\_TLLM \\
TrustLLM & Pareto-Optimality - Preference Sample & 12/21 & 1-1-2\_TLLM \\
TrustLLM & Pareto-Optimality - Pareto-Optimality & 21/21 & 1-1-3\_TLLM \\
DecodingTrust & Pareto-Optimality & 7/8 & 1-1-4\_DT \\ \hline
\end{tabular}
\end{table*}

Table VIII shows that none of the configurations in either dataset aggregate to a single model with Pareto optimality. Simply put, this shows that these sets of candidate models have tradeoffs between characteristics and that no single model dominates all others. For the TrustLLM dataset, Pareto optimality was more successful as an aggregation strategy when only used on a single level and averaging or preference sampling was used in lower levels. However, for the DecodingTrust dataset there was little aggregation, as alpaca-native was the only model removed.

\subsubsection{Pareto Optimality is a Poor Litmus Test for Trustworthiness}

Even though Pareto optimality fails as a consistent aggregation function, one might think it is a reasonable standard to identify the most trustworthy models. This is not the case. In our experimentation we have found that about half of the Pareto optimal models are not preference real—they are never chosen as the most optimal model for a preference vector even after sampling a massive number of preference vectors. This makes membership in the Pareto optimal a poor standard for trustworthiness since it often includes a large number of models that would never be any user's optimal model.
    
\begin{figure}
    \centering
    \includegraphics[scale=0.5]{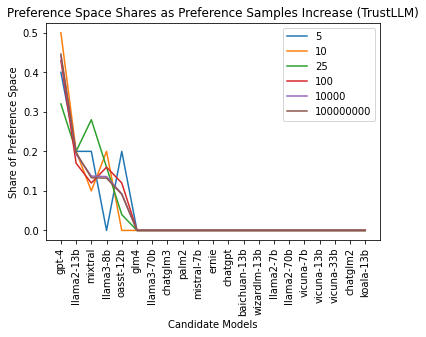}
    \caption{Shares of TrustLLM preference space as samples increase}
\end{figure}

\begin{figure}
    \centering
    \includegraphics[scale=0.5]{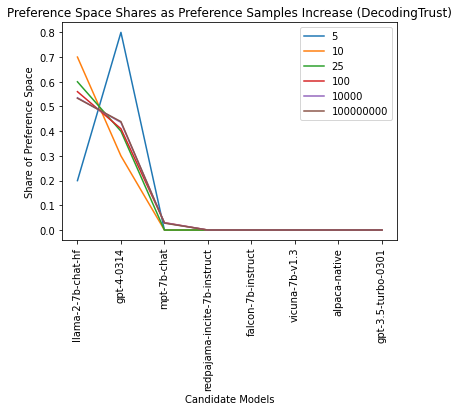}
    \caption{Shares of DecodingTrust preference space as samples increase}
\end{figure}

Figures 5 and 6 demonstrate this phenomenon. We ran symmetric, low-confidence preference sampling on both TrustLLM and DecodingTrust. For TrustLLM, we ran 10,000 samples on the bottom layer and varying sample sizes on the top layer. For DecodingTrust we ran varying sample sizes on the top layer. Because we also used 10,000 samples on the bottom level when assessing Pareto optimality in (experiment 1.1.1), we know that there are 11 Pareto optimal models for TrustLLM and that there are 7 Pareto optimal models for DecodingTrust. Even after running up to 100,000,000 preference samples, only 5/11 of TrustLLM’s Pareto optimal models were ever chosen as the optimal model according to preference sampling. For DecodingTrust, only 4/7 were ever chosen. This is somewhat surprising, because as samples tend towards infinity the preference real set converges to the Pareto set and because we sampled a huge number of preferences.

These Pareto optimal but not preference real models shouldn’t be classified as trustworthy since only an exceptionally narrow set of preferences would suggest that they are the optimal model.

\subsection{Preference Sampling Outperforms Averaging}

The main reason that averaging is an inferior aggregation function to preference sampling is that it is not sensitive to user preferences. The next sections will showcase preference sampling’s interface for asymmetric preferences and confidence as well as outline why this interface is important. Finally we will show why preference sampling improves on averaging even when a user has a low-confidence, neutral prior.	

\subsubsection{Asymmetric Preferences Cannot Be Expressed with Averaging}

\begin{table*}[htbp]
\centering
\caption{Preference sampling is sensitive to diverse user preferences}
\label{PO_aggregation}
\begin{tabular}{cccc}\hline
Aggregation Strategy & TrustLLM Optimal Model & DecodingTrust Optimal Model & Experiment ID \\ \hline
Averaging & GPT-4, 80.6\% & Llama-2, 71.5\% & 2-1 \\
Robustness Focus & GPT-4, 94.0\% & GPT-4, 81.2\% & 2-2-1 \\
Privacy Focus & Llama-2, 87.7\% & Llama-2, 91.5\% & 2-2-2 \\
Fairness Focus & Open-Assistant, 74.0\% & Llama-2, 94.1\% & 2-2-3 \\
Robustness and Privacy Focus & GPT-4, 59.8\% & Llama-2, 50.7\% & 2-2-4 \\
Robustness and Fairness Focus & GPT-4, 65.6\% & Llama-2, 66.7\% & 2-2-5 \\
Privacy and Fairness Focus & Llama-2, 81.1\% & Llama-2, 99.3\% & 2-2-6 \\
Privacy, Fairness, and Robustness Focus & Llama-2, 52.7\% & Llama-2, 89.9\% & 2-2-7 \\ \hline
\end{tabular}
\end{table*}

In experiment 2.1 we used averaging as the aggregation for every node in TrustLLM and DecodingTrust. For TrustLLM the optimal model was GPT-4 with a score of 80.6\%, for DecodingTrust the optimal model was Llama-2 with a score of 71.5\%. Unfortunately, these results are not useful to a user which values any one characteristic over another characteristic. Additionally, averaging doesn’t allow the user to specify how confident they are in their relative preferences between characteristics. Likewise, if the initial result is somewhat inconclusive—as is the case where the two frameworks disagree—there is no way to iterate on the result to understand which of the two models should be adopted.

To evaluate how trustworthiness scores change in response to non-uniform user preferences, we replaced averaging with asymmetric preference sampling that expressed a robustness-oriented preference (2-2-1), a privacy-oriented preference (2-2-2), and a fairness-oriented preference (2-2-3). As a result, we saw that GPT-4 and Llama-2 had a robustness-privacy tradeoff—both frameworks selected GPT-4 when using a robustness-oriented prior whereas both frameworks selected Llama-2 when using a privacy-oriented prior. The ontologies produced different answers for the fairness focus, but this is likely due to the difference in candidate sets tested for the two ontologies—Open Assistant appears in TrustLLM’s candidate set but not in DecodingTrust’s candidate set. Some might say that preference sampling isn’t needed for this type of analysis, as one could simply choose the best model based on a single characteristic while ignoring all other characteristics. To address this concern, we also show four more preferences in experiments 2-2-4, 2-2-5, 2-2-6, and 2-2-7 which are a combination of the earlier preferences.

In the multi-characteristic preference experiments we find that preference sampling successfully scalarizes complex preferences. For example, DecodingTrust ranks Llama-2 as very private, solidly fair, and only somewhat robust. As such, we see that the privacy and fairness co-preference (2-2-6) overwhelming endorses Llama-2 (99.3\%),  the robustness and privacy co-preference (2-2-4) solidly endorses Llama-2 (66.7\%), whereas the robustness and fairness co-preference (2-2-5) just barely endorses Llama-2 (50.7\%). When factoring in all three preferences (2-2-7), the result is somewhere between the three co-preferences (89.9\%). Though TrustLLM has a relatively higher scoring for GPT-4 than DecodingTrust, the overall trend is the same. The reason that TrustLLM ascribes a higher relative score for GPT-4 as opposed to DecodingTrust may be because DecodingTrust also evaluated GPT-3.5, which may be “taking preference” from GPT-4. This phenomenon underscores the comparative nature of preference sampling, which we will revisit in section 5D.

In addition to exposing an interface for prior knowledge to the user, preference sampling improves on averaging because of its potential as an interactive decision-making tool. It is extremely fast to conduct preference sampling due to the simplicity of sampling the Dirichlet distribution and scalarizing scores. This could allow users the ability to explore tradeoffs using preference sampling by iteratively running preference sampling with different alpha values. 

Finally, we revisit the topic of score interpretability. Where scores from averaging vary in meaning from ontology to ontology, preference sampling scores have a clear interpretation: a model’s trustworthiness score corresponds to the percentage of preference samples where it was the optimal model out of all candidate models. 

\subsubsection{Averaging Does Not Incorporate Confidence}

Another benefit that preference sampling provides is that it allows the user to quantify their confidence in the relative weights between characteristics. A simple case of this is when a user has one characteristic that they value above all others. In this case, a very confident user would provide an alpha vector where that characteristic’s value is much higher than the rest of the values in the vector, whereas a cautious user would only differentiate the characteristic slightly.

In experiments 2-3-1 through 2-3-6, we tested the effect of increasing confidence in a single-preference perspective. The result is that the trustworthiness score for the model which performed best in that category (Llama-2 for both TrustLLM and DecodingTrust) monotonically increased to near 100

\begin{table*}[htbp]
\centering
\caption{The effect of increasing confidence for a single-characteristic preference}
\label{PO_aggregation}
\begin{tabular}{ccccc}\hline
TrustLLM Optimal Model & TrustLLM Alpha Vector & DecodingTrust Optimal Model & DecodingTrust Alpha Vector & Experiment ID \\ \hline
GPT-4, 44.8\% & [1,1,1,1,1,1] & Llama-2, 56.3\% & [1,1,1,1,1,1] & 2-3-1 \\
Llama-2, 44.0\% & [1,1,1,1,2,1] & Llama-2, 68.6\% & [1,1,1,1,1,1] & 2-3-2 \\
Llama-2, 64.4\% & [1,1,1,1,3,1] & Llama-2, 79.2\% & [1,1,1,1,1,1] & 2-3-3 \\
Llama-2, 79.6\% & [1,1,1,1,4,1] & Llama-2, 85.9\% & [1,1,1,1,1,1] & 2-3-4 \\ 
Llama-2, 87.4\% & [1,1,1,1,5,1] & Llama-2, 91.4\% & [1,1,1,1,1,1] & 2-3-5 \\
Llama-2, 99.3\% & [1,1,1,1,10,1] & Llama-2, 99.4\% & [1,1,1,1,1,1] & 2-3-6 \\ \hline
\end{tabular}
\end{table*}

\begin{table*}[htbp]
\centering
\caption{Results from averaging TrustLLM ontology}
\label{avg_tllm}
\begin{tabular}{|c|*{7}{p{1.65cm}|}}\hline
Model & Trustworthiness & Truthfulness & Safety & Fairness & Robustness & Privacy & Machine Ethics \\ \hline
gpt-4 & \textbf{80.6\%} & \textbf{80.7\%} & 61.5\% & 51.4\% & \textbf{98.9\%} & 54.9\% & 69.5\% \\
ernie & 75.1\% & 66.5\% & 69.3\% & 42.0\% & 72.7\% & 76.1\% & 70.1\% \\
llama2-13b & 71.2\% & 47.1\% & 58.3\% & 51.9\% & 71.5\% & \textbf{84.1\%} & 67.4\% \\
chatgpt & 65.6\% & 66.2\% & 56.2\% & 43.8\% & 79.8\% & 48.5\% & 68.3\% \\ 
llama2-70b & 65.4\% & 48.9\% & 58.6\% & 43.2\% & 79.7\% & 61.4\% & 70.9\% \\
mixtral & 65.3\% & 71.3\% & 39.4\% & 44.9\% & 60.6\% & 55.3\% & \textbf{88.9\%} \\
glm4 & 63.3\% & 52.4\% & 47.4\% & 43.9\% & 68.9\% & 54.6\% & 87.4\% \\
wizardlm-13b & 61.7\% & 41.8\% & 67.0\% & 44.1\% & 69.6\% & 54.5\% & 69.6\% \\
vicuna-33b & 61.5\% & 48.0\% & 60.9\% & 50.2\% & 68.7\% & 45.4\% & 70.5\% \\
mistral-7b & 60.9\% & 54.8\% & 36.9\% & 57.1\% & 67.6\% & 55.1\% & 69.6\% \\
llama3-8b & 60.2\% & 53.3\% & \textbf{70.8\%} & 49.2\% & 46.6\% & 49.8\% & 65.7\% \\
llama3-70b & 56.4\% & 54.1\% & 53.1\% & 47.4\% & 48.5\% & 54.4\% & 66.5\% \\
llama2-7b & 55.3\% & 36.5\% & 57.9\% & 39.4\% & 68.8\% & 57.5\% & 65.9\% \\
vicuna-13b & 55.3\% & 38.6\% & 53.6\% & 48.8\% & 68.9\% & 51.4\% & 61.0\% \\
chatglm2 & 47.4\% & 32.1\% & 57.6\% & 33.9\% & 67.7\% & 48.6\% & 58.3\% \\
vicuna-7b & 41.1\% & 27.9\% & 42.2\% & 48.7\% & 51.9\% & 51.0\% & 47.4\% \\
oasst-12b & 40.9\% & 21.3\% & 56.6\% & \textbf{61.5\%} & 62.0\% & 35.7\% & 26.1\% \\
palm2 & 40.1\% & 27.9\% & 25.9\% & 50.1\% & 70.9\% & 32.7\% & 61.1\% \\
koala-13b & 37.1\% & 25.8\% & 60.1\% & 36.1\% & 46.2\% & 38.4\% & 49.5\% \\
baichuan-13b & 14.5\% & 33.2\% & 17.6\% & 16.5\% & 37.7\% & 28.2\% & 49.3\% \\
chatglm3 & 12.9\% & 30.7\% & 14.6\% & 26.7\% & 27.8\% & 21.3\% & 50.3\% \\ \hline
\end{tabular}
\end{table*}

\begin{table*}[htbp]
\centering
\caption{Results from preference sampling on the TrustLLM ontology}
\label{ps_tllm}
\begin{tabular}{|c|*{7}{p{1.65cm}|}}\hline
Model & Trustworthiness & Truthfulness & Safety & Fairness & Robustness & Privacy & Machine Ethics \\ \hline
gpt-4 & \textbf{44.9\%} & \textbf{69.7\%} & 1.8\% & 3.6\% & \textbf{99.7\%} & 2.5\% & 0.0\% \\
llama2-13b & 18.9\% & 0.0\% & 1.7\% & 23.5\% & 0.0\% & \textbf{64.6\%} & 0.0\% \\
llama3-8b & 13.2\% & 0.0\% & \textbf{43.1\%} & 0.0\% & 0.0\% & 0.9\% & 12.7\% \\
mixtral & 13.1\% & 11.8\% & 0.0\% & 1.5\% & 0.0\% & 0.1\% & \textbf{84.9\%} \\
oasst-12b & 9.8\% & 0.0\% & 0.0\% & \textbf{50.5\%} & 0.0\% & 0.0\% & 0.0\% \\
chatgpt & 0.0\% & 1.6\% & 0.0\% & 0.0\% & 0.0\% & 0.3\% & 0.0\% \\
chatglm2 & 0.0\% & 0.0\% & 0.0\% & 0.0\% & 0.0\% & 0.0\% & 0.0\% \\
vicuna-33b & 0.0\% & 0.0\% & 25.8\% & 2.3\% & 0.0\% & 0.0\% & 0.0\% \\
vicuna-13b & 0.0\% & 0.0\% & 0.0\% & 0.8\% & 0.0\% & 0.0\% & 0.0\% \\
vicuna-7b & 0.0\% & 0.0\% & 0.0\% & 0.0\% & 0.0\% & 0.0\% & 0.0\% \\
llama2-70b & 0.0\% & 0.0\% & 4.7\% & 0.0\% & 0.0\% & 0.7\% & 0.0\% \\
llama2-7b & 0.0\% & 0.0\% & 0.0\% & 0.0\% & 0.0\% & 0.0\% & 0.0\% \\
wizardlm-13b & 0.0\% & 0.0\% & 12.1\% & 0.0\% & 0.0\% & 0.0\% & 0.0\% \\
koala-13b & 0.0\% & 0.0\% & 0.0\% & 0.0\% & 0.0\% & 0.0\% & 0.0\% \\
baichuan-13b & 0.0\% & 0.0\% & 0.0\% & 0.0\% & 0.0\% & 0.0\% & 0.0\% \\
ernie & 0.0\% & 16.2\% & 10.7\% & 0.7\% & 0.0\% & 30.9\% & 0.0\% \\
mistral-7b & 0.0\% & 0.0\% & 0.0\% & 1.1\% & 0.0\% & 0.0\% & 0.0\% \\
palm2 & 0.0\% & 0.0\% & 0.0\% & 3.5\% & 0.3\% & 0.0\% & 0.0\% \\
chatglm3 & 0.0\% & 0.0\% & 0.0\% & 0.0\% & 0.0\% & 0.0\% & 0.0\% \\
llama3-70b & 0.0\% & 0.8\% & 0.0\% & 12.4\% & 0.0\% & 0.0\% & 0.9\% \\
glm4 & 0.0\% & 0.0\% & 0.0\% & 0.0\% & 0.0\% & 0.0\% & 1.5\% \\ \hline
\end{tabular}
\end{table*}

\subsubsection{The Low-Confidence, Symmetric Case}

Even in the case where a user is very unsure and does not expect any characteristic should be preferred over another (we call this the low-confidence, symmetric case) there are reasons to utilize preference sampling instead of averaging. First, comparisons between preference sampling scores may be more salient than comparisons between average scores. This is the case because preference sampling scores themselves are comparative: if model A is optimal for a sampled preference vector, model B cannot be. Thus, for a given analysis, all models’ preference scores could be interpreted as the probability of being the most trustworthy model in the alpha parameter's local region of the preference space.

Second, because the preference scores for a set of candidate models are a stochastic vector and because of the preference real-Pareto optimality disparity, the distribution of preference scores for a set of candidate models is sparse. For example, table XII shows the results from a preference sampling aggregation in the low-confidence, symmetric case for the TrustLLM ontology. These results show that only 5 of the 21 candidate models could feasibly be the most trustworthy model. Likewise, because the analysis is hierarchical each of the columns has this attribute of sparsity. This attribute is appealing because it narrows the set of models that a user must compare considerably.  

\section{Future Directions}

\subsection{Real-Time Multi-Objective Optimization}

This work provides a tool for the post-hoc analysis of trained models. If an AI practitioner would like to encode trustworthiness into their model from the beginning of the development process a different set of methods is needed. Several works have implemented real-time optimization of multiple objectives \cite{16_pham2024fairness} or factored constraints into their training algorithms \cite{15_hsu2018monas}. Other works which have focused on LLMs specifically use multi-objective optimization to team LLMs \cite{20_li2024s} or incorporate diverse user preferences as more fine-grained improvement to scalar rewards in reinforcement learning with human feedback (RLHF) \cite{21_wang-etal-2024-arithmetic}.

\subsection{Alternative Rankings Post-Scalarization}

As discussed in section 3A, how optimal a model is given a specific preference vector could be considered relative. For example, some preferences may scalarize models such that the gap between the 1st and 2nd most optimal models is very small while other preferences may lead to a wider gap. Our current strategy—always reward the optimal model with the same score—does not distinguish between these cases and may even disincentivize models which are frequent 2nd place finishers. Further research must be conducted to determine how crafting a reasonable dominance function can provide more desirable trustworthiness scores.

\subsection{Careful Study of Characteristics}

Several works have identified the need for more nuanced study of the interactions between characteristics of trustworthiness \cite{5_li2023trustworthy} or between characteristics and social/organizational behavior \cite{3_NIST_AI_RMF}. The tools presented in this work are only as effective as the ontologies they model, so improving our understanding of compositional trustworthiness is paramount. Finally, our work could be applied with a narrower focus on any of the characteristics of trustworthiness themselves, particularly explainability given the large amount of taxonomical work on the subject \cite{22_nauta2023anecdotal,23_vilone2020explainable}.

\subsection{The Comparative Nature of Trustworthiness}
\label{}

Because our definition of trustworthiness is fundamentally comparative, it is worth exploring how the sset of models under consideration effects the calculation of the trustworthiness score. For instance, it would be reasonable to expect that as the set of models under consideration grows, inserting or removing a single model should impact trustworthiness scores less than changes to a smaller set of models under consideration. Since from our perspective trustworthiness is a zero-sum game, voting theory could also prove useful in understanding how the freed trustworthiness from a model being removed from consideration is reallocated to other models.

\subsection{Other AI Domains}

LLMs served as an easy proving ground in which to evaluate the usefulness of preference sampling because of the availability of ontologies like TrustLLM and DecodingTrust. However, preference sampling could also be useful in other domains within AI like Visual Language Models (VLMs) which have many, potentially orthogonal, attributes and may incur tradeoffs.

\section{Conclusion}

As artificial intelligence models become increasingly more complex, trustworthiness is being assessed as a multi-criteria phenomenon. New multi-criteria assessments are descriptive and realistic, but their complexity could be a barrier for users looking to incorporate trustworthy AI research. This work addresses this need by introducing preference sampling, a method to extract a scalar trustworthiness score from multi-dimensional evaluation results by considering the many characteristics of model performance which users value. We evaluate this tool against two of the leading multi-criteria assessments of LLMs, TrustLLM and DecodingTrust, and find that preference sampling outperforms alternative aggregation functions like Pareto optimality and averaging.

\section{Licenses}

Images from DecodingTrust shared unchanged in accordance with the CC BY-SA 4.0 license.

\copyright 2024 The MITRE Corporation All Rights Reserved

\end{document}